\documentclass[nohyper,12pt,letterpaper]{JHEP3}
\usepackage{graphicx,epsfig,youngtab}

\author{Robert de Mello Koch$^2,$ Antal Jevicki$^1$, Jo\~ao P. Rodrigues$^2$ and Junggi Yoon$^{1}$\\
$^{1}$ Department of Physics, Brown University,\\
Providence, RI 02912, USA
\\
\\
$^{2}$ National Institute for Theoretical Physics,\\
School of Physics and Centre for Theoretical Physics,\\ 
University of the Witwatersrand,\\ 
Wits, 2050, South Africa\\
\qquad\\
E-mail: \email{robert@neo.phys.wits.ac.za, antal\_jevicki@brown.edu,  joao.rodrigues@wits.ac.za, jung-gi\_yoon@brown.edu}}

\abstract{
Employing the world line spinning particle picture. We discuss the appearance of several different `gauges' which we use to gain a deeper explanation of the Collective/Gravity identification. We discuss transformations and algebraic equivalences between them. For a bulk identification we develop a `gauge independent' representation where all gauge constraints are eliminated. This `gauge reduction' of Higher Spin Gravity demonstrates that the physical content of  $4$D AdS HS theory is represented by the dynamics of an unconstrained scalar field in $6$d. It is in this gauge reduced form that HS Theory can be seen to be equivalent to a $3+3$ dimensional bi-local collective representation of CFT$_3$.
}

\preprint{BROWN-HET-1658\\ WITS-CTP-138}

\title{Holography as a Gauge Phenomenon in Higher Spin Duality}


\begin{document}

\section{Introduction}

Holography, as phenomena with emergent extra dimensions  is most clearly implemented in the AdS/CFT correspondence relating $d$-dimensional QFT's to $d+1$ dimensional Gravity/and String theories. The most extensively studied example is ${\cal N}$=4 SuperYang-Mills theory and its relation to AdS$_5\times$S$^5$ String theory with features of integrability \cite{Maldacena:1997re,Gubser:1998bc,Witten:1998qj,Beisert:2010jr}.

Higher Spin equations were studied long ago \cite{Rarita:1941mf,Fronsdal:1978rb,Fang:1978wz}. Interacting theories of all spins containing Gravity were successfully constructed through a gauge principle \cite{Fradkin:1986ka,Vasiliev:1990en,Vasiliev:1995dn,Bekaert:2005vh}. Their correspondence with $N$-component vector field theories represents a significant example \cite{Klebanov:2002ja,Sezgin:2002rt} of AdS/CFT duality characterized by relative simplicity. It provides a great laboratory for studying \cite{Giombi:2009wh} and understanding some of the basic questions regarding the origin of holography and of emergent space-time.

Holography, in QFT and Gravity has been understood in a number of different schemes. In the concrete example of AdS$_{d+1}$/CFT$_d$ the origin of the extra AdS coordinate has been attributed most commonly to a `renormalization group scale'. There are other physical ideas on the emergence of space-time and Gravity. The collective field \cite{Jevicki:1980zg} approach provides a direct construct of the emergent theory as large $N$ collective phenomena. Here one has an effective re(summation) of Feynman diagrams into effective interaction vertices ( with extra dimensions).  As such the approach gives a systematic scheme for construction of the dual AdS theory and provides considerable insight into the origin of holography \cite{Das:2003vw}. In the case of $N$-component vector models, this has been demonstrated \cite{Koch:2010cy,Jevicki:2011ss,Jevicki:2014mfa} in the light-cone gauge \cite{Metsaev:1999ui} (where HS gravity is the simplest). A similar identification of AdS space in  light-cone QCD was developed in \cite{Brodsky:2013dca,Brodsky:2014yha}. In addition a renormalization group method for bi-local observables is being developed in \cite{Leigh:2014tza,Leigh:2014qca, Zayas:2013qda,Douglas:2010rc}. The collective method is easily formulated in any time-like frame, and it has been used for example in  in \cite{deMelloKoch:2012vc,Das:2012dt}. A recent  overview of the  canonical formulation of the AdS/CFT map can be found in \cite{Canonical}. A covariant version is also possible and is seen to explain \cite{Jevicki:2014mfa} some of the interesting one-loop results found in \cite{Giombi:2013fka,Giombi:2014iua}.
More complex Higher Spin correspondences involving conformal theories in 2d 
are also of high interest\cite{Gaberdiel:2012uj, Jevicki:2013kma}.
In this paper we aim to strengthen the Collective/Gravity identification and provide deeper understanding of it. We will study HS gravity through a world line spinning particle picture and will identify several different `gauges' of the theory. The most well known one, which we call the Fronsdal gauge, translates in the second-quantized version into the Higher Spin equations of Fronsdal with de Donder gauge condition. We then introduce a more symmetric or `bi-local' gauge which features space-time and internal coordinates in a symmetrical way. It is in this gauge that the appearance of CFT in its collective representation becomes manifest. We also discuss the transformations and algebraic equivalences between the different gauges.

For a bulk identification between the two sides it is useful to have Gravity (and HS theory) written in the `gauge independent' representation where all gauge constraints are solved. This `gauge reduction' can be quite nontrivial in any gauge theory and we perform it in Higher Spin Gravity. 
We demonstrate in particular that the physical content of $4$D AdS HS theory is represented by a single scalar field 
dynamics in $6$d. 
Since in this (physical) version one has no gauge conditions or redundant variables, consequently the identification with $3+3$ dimensional bi-local collective equations becomes visible.
This demonstration is performed at the quadratic level, which determines the spectrum of the theory, 
One can hope that the features and reductions identified extend to the case of interactions.

The outline of the paper is as follows: 
In Chapter 2 we give the description of the world line particle formulation of higher spins and write down the different gauges that we consider. Chapter 3 discuss the appearance of the bi-local collective equations in the symmetric gauge. Chapter 4 presents the reductions to unconstrained fields, and the process of solving the gauge conditions demonstrating reduction to a 6d scalar field  equation. Equivalence relations between the gauges are demonstrated in Chapter 5. In the Conclusions, Chapter 6, we give a summary of our demonstration.

\section{Gauges}
Let us begin by discussing Higher Spin field equations in terms of a world line (first quantized) particle in AdS$_{d+1}$. It is useful to embed AdS$_{d+1}$ into a $d+2$ dimensional space-time $R^{d,2}$, since the $SO(d,2)$ conformal symmetry is manifestly realized \cite{Fronsdal:1978vb}. Even though some of the kinematical facts will hold for any dimension $d$ we will concentrate specifically on the  3d case representing AdS$_4$/CFT$_3$ duality.
In terms of the coordinates $X^A$ of the embedding space and the coordinates $x^\mu$ of the AdS space, the higher spin field $h_{\mu_1\cdots\mu_s}(x)$ is related to a higher spin field in embedding space as
\begin{equation}
H_{A_1\cdots A_s}(X)=x^{\mu_1}_{,A_1}\cdots x^{\mu_s}_{,A_s}h_{\mu_1\cdots\mu_s}(x)
\end{equation} 
where $x^{\mu}_{,A}=\partial x^\mu/\partial X^A$.
To generate all spins one introduces the internal(spin) coordinate as a copy of $R^{d,2}$ denoted by $Y^A$.
The higher spin field is then
\begin{equation}
  H(X,Y)=\sum_s H_{A_1\cdots A_s}(X) Y^{A_1}\cdots Y^{A_s}
\end{equation}
For the specific case of AdS$_4$ which we focus on  we have $SO(3,2)$ realized on the tensor product of two copies of $R^{3,2}$.
Introducing the momenta ($\eta_{AB}=(-,-,+,+,+)$)
\begin{equation}
 P_A=-i{\partial\over\partial X^A}\qquad K_A=-i{\partial\over\partial Y^A}
\end{equation}
conjugate to $X^A$ and $Y^A$ respectively.
The generators are
\begin{equation}
L_{AB} = P_A X_B - P_B X_A + K_A Y_B - K_B Y_A
\end{equation}
For the  massless spin $s$ theory, which is to be associated with the $D(s+1,s)$ representations of $SO(3,2)$, one can constrain the second and fourth order Casimir operators as
\begin{equation}
C_2+E_0^2+s^2=0\quad,\quad C_4+E_0^2 s^2=0
\end{equation}
It will be useful to write out the explicit forms 
\begin{eqnarray}
   C_2&=&\frac{1}{2}L_{AB}L^{AB}\cr
         &=&X^2 P^2 -(X\cdot P)^2+Y^2 K^2-(Y\cdot K)^2+2X\cdot Y P\cdot K -2X\cdot KY\cdot P 
\end{eqnarray}
\begin{eqnarray}
    C_4&=&\frac{1}{4}L_{AB}L^B{}_CL^C{}_DL^{DA}-\frac{1}{2}\left(\frac{1}{2}L_{AB}L^{AB}\right)^2\cr
          &=&X^2\left(K^2(Y\cdot P)^2+P^2(Y\cdot K)^2-2(P\cdot K)(Y\cdot P)(Y\cdot K)\right)\cr
          &&+Y^2\left(K^2 (X\cdot P)^2+P^2(X\cdot K)^2-2(K\cdot P)(X\cdot P)(X\cdot K)\right)\cr
          &&+(X^2Y^2-(X\cdot Y)^2)\left( (P\cdot K)^2-P^2K^2\right)-(X\cdot K)^2(Y\cdot P)^2\cr
          &&-(X\cdot P)^2(Y\cdot K)^2+2(X\cdot P)(X\cdot K)(P\cdot Y)(Y\cdot K)\cr
          &&+2(P\cdot K)(X\cdot P)(X\cdot Y)(Y\cdot K)+2(P\cdot K)(X\cdot K)(X\cdot Y)(Y\cdot P)\cr
          &&-2 P^2 (X\cdot K)(X\cdot Y)(Y\cdot K)-2K^2(X\cdot P)(X\cdot Y)(Y\cdot P)
\end{eqnarray}
Eliminating $s$ one has the following equation(constraint):
\begin{equation}
   L=C_4+{1\over 4}C_2^2 = 0
\end{equation}
This equation for the spinning particle is analogous to the Laplacian constraint.

This restriction of the Casimirs, however, is still not sufficient to specify the irreducibility of representations, and one needs to impose further  first class constraints. These are not unique and their specification corresponds to different gauges of the theory \cite{Jevicki:2011ss}. Several cases starting with the Fronsdal's one will be  given below.

\subsection{Fronsdal /de Donder Gauge}

The gauge invariant equation of motion for a symmetric traceless $s$-tensor gauge field $h_{\mu_1\cdots\mu_s}$ in AdS$_{d+1}$ is given by
\begin{equation}
  (\Box-m^2)h_{\mu_1\cdots\mu_s}+s\nabla_{(\mu_1}\nabla^\nu h_{\mu_2\cdots\mu_s)\nu}-
   {s(s-1)\over 2(d+2s-3)}g_{(\mu_1\mu_2}\nabla^{\nu_1}\nabla^{\nu_2}h_{\mu_3\cdots\mu_s)\nu_1\nu_2}=0
   \label{waveeqn}
\end{equation}
where gauge symmetry fixes
$$
    m^2=(s-2)(d + s-3)-2
$$
In his original treatment of higher spin fields, Fronsdal \cite{Fronsdal:1978vb} employed a covariant gauge specified by
\begin{equation}
   \nabla^\rho h_{\rho\mu_2\cdots \mu_s}=0\qquad g^{\rho\sigma}h_{\rho\sigma\mu_3\cdots \mu_s}=0
\end{equation}
In this gauge, the equation of motion becomes
\begin{equation}
  (\Box-m^2)h_{\mu_1\cdots\mu_s}=0
\end{equation}
To have transversality and tracelessness one imposes, following Fronsdal \cite{Fronsdal:1978vb}, the following four constraints
\begin{equation}
T_1 = X^A P_A + Y^A K_A + 1 = 0
\end{equation}
\begin{equation}
T_2 = X^A K_A = 0
\end{equation}
\begin{equation}
T_3 = K^A K_A = 0
\end{equation}
\begin{equation}
T_4 = P^A K_A = 0
\end{equation}
These first class constraints specify the Fronsdal gauge. In the phase space one is free to add to these
certain gauge conditions. Specifically we will make use of the ``gauge conditions''
\begin{equation}
T_{-1} = X^A X_A +r^2 = 0\label{AdSurf}
\end{equation}
\begin{equation}
T_{-2} = X^A Y_A = 0
\end{equation}
where $r$ is the radius of the AdS spacetime.

With the above constraints we find that
\begin{equation}
   L=C_4+{1\over 4}C_2^2 = \frac{1}{4}\left(P^2\right)^2\label{Laplace}
\end{equation}
which represents the Laplace operator in the Fronsdal gauge.
To explicitly verify that (\ref{Laplace}) reproduces (\ref{waveeqn}), change variables in the
embedding space into the radial coordinate $r$ and four coordinates that parametrize the constraint 
surface determined by (\ref{AdSurf}). 
Restricting the embedding space Laplacian to the constraint surface and projecting tensors to the space 
tangent to the constraint surface then reproduces (\ref{waveeqn}).

\subsection{KLSS Gauge}

In \cite{Kuzenko:1994vh,Kuzenko:1995aq}  another description of the spinning Anti-de Sitter particle was given and it was shown to reduce to a description on AdS$_4$ times $S^2$. The four first class constraints of \cite{Kuzenko:1994vh,Kuzenko:1995aq} are
\begin{equation}
\frac{1}{2} (-X\cdot P_X + Y\cdot P_Y) =0
\end{equation}
\begin{equation}
\frac{1}{2}( X\cdot P_X + Y\cdot P_Y) =0
\end{equation}
\begin{equation}
P_Y \cdot P_Y=0
\end{equation}
\begin{equation}
X \cdot P_Y=0
\end{equation}
We refer to this as KLSS Gauge. Again one adds ``gauge conditions'' which can be identical ones in the Fronsdal case.

\subsection{Symmetric Gauge}

Finally we give another even more symmetric gauge, which will turn out to be related to the bi-local collective field description.
The bilocal fields reside in $2+1$ dimensional Minkowski spacetime, with events labeled by $U$ and $V$.

One writes the following four first class constraints
\begin{equation}
\frac{1}{2} V\cdot P_V=0
\end{equation}
\begin{equation}
\frac{1}{2} U\cdot P_U =0
\end{equation}
\begin{equation}
U \cdot U =0
\end{equation}
\begin{equation}
V \cdot V =0
\end{equation}
A single ``gauge condition" $U\cdot V=1$ will be of relevance.

We will discuss this gauge in much more detail in the following section, as it represents the bi-local/collective field version of the theory. We will also demonstrate an algebraic equivalence between the various gauges specified above. We mentioned that in the present work we are using the simplest version of spinning particle dynamics. There are a number of other relevant studies of spinning particles in AdS space, in particular \cite{Howe:1988ft,Bastianelli:2008nm}. It will be interesting to incorporate the present scheme in future work as part of the more general a tensor particle theory \cite{Bandos:1999qf,Bandos:1999pq,Didenko:2003aa} and a so called ``parent theory'' \cite{Barnich:2006pc} from which Fronsdal's and Vasiliev's unfolded formulations are known to follow through two different reductions.

\section{Collective Field /Symmetric Gauge}

Collective field theory of the $O(N)$ vector model concentrates on the dynamics of the composite, bi-local field
\begin{equation}
  \Psi (x_1^\mu,x_2^\mu)=\varphi(x_1)\cdot\varphi(x_2)
\end{equation}
which takes place in the six dimensional space given by the tensor product of two copies of $R^{2,1}$. In this section the index $\mu=0,1,2$ is a vector index for $2+1$ Minkowski space $R^{2,1}$. Its dynamics is fully specified by the (collective) action
\begin{equation}
   S=\int d^3 x \left( -\Delta_x\Psi (x,y)\Big|_{x=y}\right)-{N\over 2} \, {\rm Tr}\log\Psi
\end{equation}
which is directly deduced from the QFT. The QFT Lagrangian represents the first term in the above expression. The second term encapsulates all the quantum effects, and $N$ which now appears as a coupling constant defines the complete nonlinearity in this collective representation.

After a shift by the stationary background 
\begin{equation}
   \Psi(x_1,x_2) =\Psi_0(x_1,x_2)+\tilde\Psi (x_1,x_2)
\end{equation}
one gets the linearized equations and a sequence of $1/N$ vertices:
\begin{equation}
   \partial_1^2\partial_2^2\tilde\Psi (x_1,x_2)+\sum_{n=3}^\infty N^{1-{n\over 2}}n\int
\prod_{l=1}^{n-2} d^3 y_ l {\partial\over\partial y^l}{\partial\over\partial y^l}
\partial_1^2\partial_2^2 
\tilde\Psi (x_1,y_1)\tilde\Psi (y_1,y_2)\cdots \tilde\Psi (y_{n-2},x_2)= 0
\end{equation}
It was proposed \cite{Das:2003vw} that this action and the associated collective equations define the gravitational dual of the $O(N)$ vector CFT. As emphasized in \cite{Das:2003vw} this represents a bulk description of Higher Spin theory. The emergent AdS space-time can be most clearly identified in the light-cone gauge \cite{Koch:2010cy}. We will, in what follows, extend this identification to the covariant case and demonstrate that the  bi-local description can be directly associated with the symmetric gauge of Higher Spin theory.

In the world line description, we consider the particle variables introduced in the previous section.
Denote the variables conjugate to $U,V$ by $P_U,P_V$. After a Fourier transform, the constraints are
\begin{equation}
\frac{1}{2}V\cdot P_V=0
\end{equation}
\begin{equation}
\frac{1}{2} U\cdot P_U =0
\end{equation}
\begin{equation}
U \cdot U =0
\end{equation}
\begin{equation}
V \cdot V =0
\end{equation}

With the above constraints we find that
\begin{equation}
   C_4+{1\over 4}C_2^2 = P_U^2 P_V^2
\label{CollCas}
\end{equation}
defining the Laplace operator in this gauge.

We have that the field $\Psi(U,V)$, is defined in the $5+5$ dimensional space obtained by taking two copies of $R^{3,2}$. To obtain the unconstrained physical description we need to solve the pairs of constraints $U\cdot P_U=0,$ $U \cdot U=0$ and $V\cdot P_V =0,$ $V\cdot V =0$. We will now demonstrate that after solving the constraints we obtain a $3+3$ dimensional bi-local description with the correct collective dynamics.

We will describe in detail the solution to $U\cdot P_U=0,$ $U\cdot U =0$. The solution to the second pair of constraints follows exactly the same logic. To solve the constraints it is useful to introduce the light cone momenta
\begin{equation}
 U^{\pm}= U^3\pm  U^5
\end{equation}
To implement the constraint $U\cdot P_U \Psi\left(U\right)=0$ where $U\cdot P_U$ generates scale transformation
\begin{equation}
U^{\pm} \to \lambda U^{\pm}\quad  U^\mu\to\lambda U^\mu
\end{equation}
we express the wave function in terms of invariants under scaling. (In this section, we use $\mu=01,2,3$ as index for three-dimensional Minkowski space) One can choose
\begin{equation}
u^\mu={U^\mu \over U^-}\quad,\quad {U^+\over U^-}
\end{equation} 
as independent set of invariants. However, because of $U\cdot U=0$, we have
\begin{equation}
 {U^+\over U^-}=-\frac{1}{2}u_\mu u^\mu
\end{equation}
so that it does not represent an independent variable. Hence, our reduced wave function reads
\begin{equation}
\Psi\left(U\right)=\Psi\left(u^\mu\right)
\end{equation}
To obtain the reduced form for the $SO\left(2,3\right)$ generators, we proceed as follows. One can express the momentum in the embedding space in terms of $U^-$ and independent invariants $u^\mu$'s according to chain rule.
\begin{eqnarray}
P_{U\mu} &\to&{\partial\over\partial U^\mu}={1\over U^-}{\partial\over\partial u^\mu}\cr
P_{U-} &\to& {\partial\over\partial U^-}=-{u^\mu\over U^-}{\partial\over\partial u^\mu}
\end{eqnarray}
This expresses $U$ and $P_U$ in terms of $u^\mu, U^-$. The answers for the generators will depend only on the invariant variables $u^\mu$. Performing the same reduction for the pair $(V,P_V)$, the collective wave function, after performing both reductions is $\Psi (u^\mu,v^\mu)$. The original $SO(3,2)$ generator
\begin{equation}
   L_{AB} = {P_U}_A U_B - {P_U}_B U_A + {P_V}_A V_B - {P_V}_B V_A
\end{equation}
become
\begin{eqnarray}
 L_{\mu +}&=&P_\mu\qquad L_{\mu-}=K_\mu\qquad L_{+-}=D\cr
P_\mu&=&{\partial \over \partial u^\mu}+{\partial \over \partial v^\mu}\cr
M_{\mu\nu}&=&-u_\mu{\partial\over\partial u^\nu}+u_\nu{\partial\over\partial u^\mu}-v_\mu{\partial\over\partial v^\nu}+v_\nu{\partial\over\partial v^\mu}\cr
D&=&u^\mu{\partial\over\partial u^\mu}+v^\mu{\partial\over\partial v^\mu}+1\cr
K_\mu&=&-{1\over2}u^2 {\partial\over\partial u^\mu}+u_\mu u^\nu{\partial\over\partial u^\nu}
-{1\over2}v^2 {\partial\over\partial v^\mu}+v_\mu v^\nu{\partial\over\partial v^\nu}
\end{eqnarray}
which are seen to be the generators of the three-dimensional conformal group acting on a bi-local field. The field in linearized approximation will obey an equation following from the reduction of the Casimirs: from eq.~(\ref{CollCas}) one indeed obtains 
\begin{equation}
C_4+\frac{1}{4}C_2^2={1\over 4} \left(u-v\right)^2 p_u^2 p_v^2 
\end{equation}
where we have $\left(u-v\right)^2$ factor in addition to eq.~(\ref{CollCas}). This factor appears because the above reduction does not satisfy the ``gauge condition'' $U\cdot V=1$ in general. A field-dependent gauge transformation to $U\cdot V=1$ gauge will eliminate $\left(u-v\right)^2$ factor in the collective Laplacian. To summarize, we have shown that the bi-local collective field equations (in leading order) can be obtained from a symmetric gauge fixing of higher spins.

\section{Gauge Reduction}

We now proceed to a direct method for demonstrating equivalence of collective and higher spin equations.
The non triviality of direct identification comes from the fact that HS fields require gauge fixing conditions, while the bi-local field of collective theory is not constrained. So one strategy for a comparison is to solve the gauge constraints imposed on the  Higher Spins and obtain equations entirely in terms of independent (gauge invariant) variables with no constraints. This is usually difficult to do. We will be able to perform this reduction in the Higher Spin case and show that it leads to a scalar field dynamics $6$ dimensions, which are split into 4 of AdS$_4$ and a $2$-sphere  S$^2$ for the reduced spin degrees of freedom. A specific example of this reduction to physical degrees of freedom was first presented in the spinning particles framework in the work of \cite{Kuzenko:1994vh,Kuzenko:1995aq} which we describe first. We will then demonstrate a that a very similar reduction holds for the Fronsdal HS case \cite{Fronsdal:1978vb}.

\subsection{KLSS reduction to AdS$_4\times$S$^2$}

In this subsection we start from two copies of the five dimensional flat space $R^{3,2}$ with coordinates and momenta $(X^A,P^A)$ and $(Y^A,K^A)$ for the two copies. The copy of $R^{3,2}$ with coordinates $Y^A$ is used to package the complete set of higher spin fields into a single field. After imposing the constraints introduced above, which implement the KLSS gauge, we are left with the $6$ dimensional
space AdS$_4\times$S$^2$.
The fields on this space are unconstrained since to obtain this description all of the gauge constraints have been solved. 
We introduce symmetric coordinates $q^m$ for the AdS$_4$ and complex coordinates $z$ for the $S^2$.
AdS$_4$ is the physical space-time while the $S^2$ is used to collect the complete set of higher spin fields into a single field.

In \cite{Kuzenko:1994vh,Kuzenko:1995aq} the constraints implementing the KLSS gauge were solved. We will briefly summarize this reduction. The constraints $X\cdot Y=0$, $X\cdot P=0$ and $X\cdot K=0$ eliminate one component from each of $Y,P,K$, by forcing them to lie within the subspace orthogonal to $X$. Within this subspace we still need to impose $Y^A K_A=0$ and $K^A K_A=0$. These constraints define the Dirac cone which was studied in detail in the last section. The momenta $\bar K^\mu$ transervse to $X$ can be written using the spinor helicity formalism as
\begin{equation}
  \bar K_{a\dot{b}}=\left[\matrix{-\bar K^0 +\bar K^3 &\bar K^1 -i\bar K^2\cr \bar K^1 +i\bar K^2 &-\bar K^0 -\bar K^3}\right]
\end{equation}
The fact that $\bar K$ is null implies that $\det \bar K_{a\dot{b}}=0$ and hence that $\bar K_{a\dot{b}}$ is the outer product of a single vector
\begin{equation}
  \bar K_{a\dot{b}}=v_a\bar{w}_{\dot{b}}
\end{equation}
The fact that $\bar K_{a\dot{b}}$ is hermitian means that $\bar{w}_{\dot{b}}=v_a^*$. Following \cite{Kuzenko:1994vh,Kuzenko:1995aq} introduce two spinors
\begin{equation}
  \omega_a = (1,-1/z)\qquad z_a=(-z,1)
\end{equation}
\begin{equation}
  \omega^a = (-1/z,-1)\qquad z^a=(1,z)
\end{equation}
The variables $z$ and $\bar z$ are the coordinates of the $S^2$. Parametrize AdS$_4$
\begin{equation}
   (X^0)^2 + (X^5)^2 - (X^1)^2 - (X^2)^2 - (X^3)^2 = r^2
\end{equation}
by the coordinates $q^\mu$ with
\begin{equation}
  X^0 = {2rq^0\over 1+q\cdot q}\qquad X^1 = {2rq^1 \over 1+q\cdot q}
\end{equation}
\begin{equation}
  X^2 = {2 r q^2\over 1 + q\cdot q}\qquad X^3 = {2 r q^3 \over 1 + q\cdot q}
\end{equation}
\begin{equation}
  X^5 ={r(1-q\cdot q)\over 1+q\cdot q}\qquad q\cdot q = (q^0)^2-(q^1)^2-(q^2)^2-(q^3)^2
\end{equation}
It is straightforward to verify that
\begin{equation}
   ds^2 =-{4r^2 dq\cdot dq\over (1+q\cdot q)^2}
\end{equation}
On AdS$_4\times S^2$, in terms of the above coordinates, the generators of the $SO(3,2)$ algebra are
($I,J=1,2,3$)
\begin{eqnarray}
L_{0I}&=&q^0 p_I + q^I p_0 +S_{0I}\cr
L_{IJ}&=&q^I p_J - q^J p_I +S_{IJ}\cr
L_{I5}&=&R p_I + {1\over 4R} (2q^I q^\mu p_\mu + q^2 p_I)+{q^\mu\over 2R}S_{I\mu}\cr
L_{05}&=&R p_0 - {1\over 4R} (2q^0 q^\mu p_\mu - q^2 p_0)+{q^\mu\over 2R}S_{0\mu}
\end{eqnarray}
with
\begin{eqnarray}
S_{\mu\nu}&=&-(\sigma_{\mu\nu})_{\alpha\beta}z^\alpha z^\beta p_z + (\bar\sigma_{\mu\nu})_{\dot\alpha\dot\beta}\bar z^{\dot\alpha} \bar z^{\dot \beta} p_{\bar z}\cr
\{ q^\mu,p_\nu\}&=&\delta^\mu_\nu\qquad \{z,p_z\}=1\qquad \{\bar z,p_{\bar z}\}=1
\end{eqnarray}

\subsection{Fronsdal case to AdS$_4\times$S$^2$}

In this subsection we again start from two copies of the five-dimensional flat space $R^{3,2}$ with coordinates and momenta $(X^A,P^A)$ and $(Y^A,K^A)$ for the two copies. The copy of $R^{3,2}$ with coordinates $Y^A$ is again used to package the complete set of higher spin fields into a single field. After imposing the constraints introduced above, which implement the Fronsdal gauge, we are left with the $6$ dimensional space AdS$_4\times$S$^2$. The constraints defining the Fronsdal and KLSS gauge are different, but the resulting physical space is the same. Since to obtain this description all of the gauge constraints have been solved, the result should be gauge invariant so that this is not unexpected. We introduce Poincare coordinates $x^\mu$ for the AdS$_4$ and coordinates $\theta,\phi$ for the $S^2$.
AdS$_4$ is the again physical space-time while the S$^2$ is again used to collect the complete set of higher spin fields into a single field.

In Fronsdal's gauge we have the four second class constraints, $T_1,T_2,T_{-1},T_{-2}$, and two first class ones $T_3, T_4$. First of all, we will solve the four second class constraints. Under a transformation,
\begin{equation}
\left(X,^AP^A,Y^A,K^A\right)\quad\longrightarrow\quad \left(X^A,P^A-\frac{X^A}{X\cdot X},Y^A,K^A\right)\label{constraint:transformation1}
\end{equation}
the ordering term in $T_1$ vanishes and other constraints are invariant up to linear combination.
The only change is in $T_1$ which becomes
\begin{equation}
T'_1=X\cdot P+Y\cdot K
\end{equation}
The second class constraints, $T'_1,T_2,T_{-1},T_{-2}$ are solved by 
\begin{eqnarray}
X^a&=&\frac{x^a}{z}\label{eq1:reduction from 5 to 4}\\
X^3&=&\frac{1}{2}\left(\frac{1}{z}-\frac{x^ax_a}{z}-z\right)\\
X^5&=&\frac{1}{2}\left(\frac{1}{z}+\frac{x^a x_a}{z}+z\right)\\
P_A&=&\frac{\partial x^\mu}{\partial X^A} p_\mu+\left(\frac{\partial^2 x^\mu}{\partial X^A\partial X^B}\frac{\partial X^B}{\partial x^\nu}\right)y^\nu k_\mu\\
Y^A&=&\frac{\partial X^A}{\partial x^\rho}y^\rho\\
K_A&=&\frac{\partial x^\nu}{\partial X^A}k_\nu\label{eq6:reduction from 5 to 4}
\end{eqnarray}
where we use the convention for indices in this section.
\begin{eqnarray}
a,b,\cdots=0,1,2,\qquad i,j,\cdots=1,2,\qquad I,J,\cdots=1,2,3,\qquad \mu,\nu,\cdots=0,1,2,3\nonumber
\end{eqnarray}
With this solution, the remaining first constraints become
\begin{eqnarray}
T_3&=&z^2 \eta_{\mu\nu}k^\mu k^\nu\\
T_4&=&zy^3 \eta_{\mu\nu}k^\mu k^\nu+z^2\eta_{\mu\nu}k^\mu p^\nu
\end{eqnarray}
where $\eta_{\mu\nu}=\mbox{diag}\left(-1,+1,+1,+1\right)$. Ignoring ordering issues, we need to solve the following two equations
\begin{eqnarray}
k^\mu k_\mu\phi\left(x;y\right)&=&0\label{traceless condition}\\
p^\mu k_\mu\phi\left(x;y\right)&=&0\label{covariant gauge condition}
\end{eqnarray}
where we used the flat metric and
\begin{equation}
\phi\left(x;y\right)=\phi_{\mu_1\cdots \mu_s}y^{\mu_1}\cdots y^{\mu_s}
\end{equation}
Note that one can get the same equations starting from the traceless condition and in covariant gauge. i.e.
\begin{eqnarray}
\eta^{\mu_1 \mu_2}\varphi_{\mu_1\mu_2\cdots\mu_s}&=&0\\
\left(\partial_z-\frac{2}{z}\right)\varphi_{z\mu_2\cdots\mu_x}+\partial_i\varphi_{i\mu_2\cdots \mu_s}&=&0
\end{eqnarray}
where $\varphi_{\mu_1\cdots\mu_s}\left(x\right)$ is Fronsdal's higher spin field. Redefining the field $\varphi$,
\begin{equation}
\phi_{\mu_1\cdots \mu_s}=\frac{1}{z^2}\varphi_{\mu_1\cdots \mu_s}
\end{equation}
one recovers eq.~(\ref{traceless condition}) and eq.~(\ref{covariant gauge condition}). In \cite{Metsaev:1999ui}, Metsaev solved the system of equations eq.~(\ref{traceless condition}) and eq.~(\ref{covariant gauge condition}). We will follow a similar procedure. Consider a Fock space which consists of
\begin{equation}
\left|\phi\left(x;y\right)\right>=\phi_{\mu_1\mu_2\cdots \mu_s}\left(x\right)\;y^{\mu_1}y^{\mu_2}\cdots y^{\mu_s}\left|0\right>
\end{equation}
The traceless condition eq.~(\ref{traceless condition}) and the covariant gauge condition eq.~(\ref{covariant gauge condition}) can be written as
\begin{eqnarray}
k^\mu k_\mu\left|\phi\right>&=&0\label{traceless condition2}\\
p^\mu k_\mu\left|\phi\right>&=&0\label{covariant gauge condition2}
\end{eqnarray}
Introduce kernels, $\mathcal{M}_1,\mathcal{M}_2,\mathcal{M}_3$ and $\mathcal{M}_4$ to manipulate 
eq.~(\ref{traceless condition2}) and eq.~(\ref{covariant gauge condition2}).
\begin{eqnarray}
\mathcal{M}_1&\equiv&\exp\left[-y^0\left(\frac{1}{p^0}k^I  p_I\right)\right]\\
\mathcal{M}_2&\equiv& \exp\left[-\theta_1\left(y^1k^2-y^2k^1\right)\right]\\
\mathcal{M}_3&\equiv& \exp\left[-\theta_2\left(y^3k^1-y^1k^3\right)\right]\\
\mathcal{M}_4&\equiv&\exp\left[-y^3k^3\log\left(\frac{1}{p^0}\sqrt{-p^\mu p_\mu}\right)\right]
\end{eqnarray}
where
\begin{eqnarray}
\theta_1\equiv\arctan\frac{p^2}{p^1}\qquad,\qquad\theta_2\equiv \arctan \frac{\sqrt{p^ip_i}}{p^3}
\end{eqnarray}
Define a new basis for the Fock space, $\left|\Phi\left(x;y^0,y^1,y^2,y^3\right)\right>$.
\begin{equation}
\left|\phi\left(x;y^0,y^1,y^2,y^3\right)\right>=\mathcal{M}_1\mathcal{M}_2\mathcal{M}_3\mathcal{M}_4 \left|\Phi\left(x;y^0,y^1,y^2,y^3\right)\right>
\end{equation}
In $\left|\Phi\left(y^0,y^1,y^2,y^3\right)\right>$ basis, eq.~(\ref{traceless condition2}) and eq.~(\ref{covariant gauge condition2}) takes the following form.
\begin{eqnarray}
\left(k^Ik_I+k^0f\left(p,k\right)\right)\left|\Phi\left(x;y^0,y^1,y^2,y^3\right)\right>&=&0\label{traceless condition3}\\
-k^0 p^0\left|\Phi\left(x;y^0,y^1,y^2,y^3\right)\right>&=&0\label{covariant gauge condition3}
\end{eqnarray}
where $f\left(p,k\right)$ is an unimportant function of $p$ and $k$. The covariant gauge condition eq.~(\ref{covariant gauge condition3}) can be easily solved by $\left|\Phi\left(x;y^1,y^2,y^3\right)\right>$ which is independent of $y^0$. In $\left|\Phi\left(y^1,y^2,y^3\right)\right>$ basis, the traceless condition reads
\begin{equation}
k^Ik_I\left|\Phi\left(y^1,y^2,y^3\right)\right>=0
\end{equation}
This traceless condition is also solved by spherical harmonics of $y^1,y^2,y^3$. i.e. when restricting on spin-$s$ field, one has
\begin{equation}
\left|\Phi^{sol}_{s,m}\left(x;y^1,y^2,y^3\right)\right>=\Phi\left(x\right) Y_{s,m}\left(y^1,y^2,y^3\right)\left|0\right>
\end{equation}
Hence, we obtain AdS$_4\times S^2$ by solving all Fronsdal constraints. Now, we will calculate the representation of $SO(2,3)$ for AdS$_4\times S^2$. We start with the $SO(2,3)$ generators for the $(5+5)$-dimensional embedding space
\begin{eqnarray}
L^{AB}&=&X^AP^B-X^BP^A+Y^AK^B-Y^BK^A\cr
&=&\left(\matrix{
{J^{ab}} &  {-\frac{1}{2}P^a-K^a} & {-\frac{1}{2}P^a+K^a}\cr
{\frac{1}{2}P^b+K^b} & {0} & {D}\cr
{ \frac{1}{2}P^b-K^b} & {-D} & {0}\cr
}\right)
\end{eqnarray}
With the solution eq.~(\ref{eq1:reduction from 5 to 4})$\sim$eq.~(\ref{eq6:reduction from 5 to 4}), one can easily obtain the form of the $SO(2,3)$ generators for $(4+4)$-dimensional space. Note that the transformation in eq.~(\ref{constraint:transformation1}) does not change the form of the generators. Finally, using $\mathcal{M}_1,\cdots,\mathcal{M}_4$, one can obtain representation of $SO(2,3)$ in the $\left|\Phi^{sol}_{s,m}\right>$ basis
\begin{eqnarray}
P^0&=&p^0\\
P^1&=&p^1\\
P^2&=&p^2\\
D&=&x^\mu p_\mu+s
\end{eqnarray}
\begin{eqnarray}
J^{01}&=&tp^1-x^1 p^0+M^{12} \frac{p^0 p^2 p^3}{(\hat{p})^2 (\bar{p})^2}-M^{23}\frac{p\, p^2}{\hat{p}\bar{p}}-M^{31}\frac{p^1p^3p}{\hat{p} (\bar{p})^2 }\\
J^{12}&=&x^1p^2-x^2p^1\\
J^{20}&=&x^2p^0-x^0p^2+M^{12} \frac{p^0 p^1 p^3}{(\hat{p})^2 (\bar{p})^2}-M^{23}\frac{p\, p^1}{\hat{p}(\bar{p})^2}+M^{31}\frac{p^2p^3p }{\hat{p} (\bar{p})^2}
\end{eqnarray}
\begin{eqnarray}
K^0&=&-\frac{1}{2}x^\mu x_\mu p^0+tD+\left(M^{12}\right)^2\frac{p^0}{2}\left(-\frac{1}{(\hat{p})^2}
+\frac{2}{(\bar{p})^2}-\frac{1}{(p)^2}\right)\cr
&&-M^{IJ}M_{IJ}\frac{p^0}{2}\left(\frac{1}{(\bar{p})^2}-\frac{1}{(p)^2}\right)+M^{12}M^{23}\frac{p^3\, p}{\hat{p}(\bar{p})^2}+M^{12}\frac{p^0p^3J^{12}}{\hat{p}^2\bar{p}}\cr
&&-M^{23}\frac{p\, J^{12}}{\hat{p}\bar{p}}+M^{31}\frac{p\left(-p^3\left(x^i p_i\right)+(\hat{p})^2 z\right)}{\hat{p}(\bar{p})^2}
\end{eqnarray}
\begin{eqnarray}
K^1&=&-\frac{1}{2}x^\mu x_\mu p^1+x^1D+\left(M^{12}\right)^2\frac{p^1}{2}\left(\frac{1}{(\hat{p})^2}-\frac{2}{(\bar{p})^2}-\frac{1}{(p)^2}\right)+M^{IJ}M_{IJ}\frac{p^1\left(p^0\right)^2}{2(\bar{p})^2 (p)^2}\cr
&&-\left(-M^{31}s+M^{12}M^{23}\right)\frac{p^0p^1p^3}{\hat{p}p (\bar{p})^2}
+\left(M^{23}s+M^{12}M^{31}\right)\frac{p^0p^2}{\hat{p}\, p\, \bar{p}} \cr
&&-M^{12}s\frac{p^2 p^3}{(\hat{p})^2 \bar{p}}
+M^{12}\frac{ p^2\left(p^0p^3t- (\bar{p})^2 z\right)}{(\hat{p})^2\bar{p}}-M^{23}\frac{p^2\, p\, }{\hat{p}\bar{p}}\cr
&&-M^{31}\frac{p^1p^3\, p\, t}{\hat{p}(\bar{p})^2}
\end{eqnarray}
\begin{eqnarray}
K^2&=&-\frac{1}{2}x^\mu x_\mu p^2+x^2D+\left(M^{12}\right)^2\frac{p^2}{2}\left(\frac{1}{(\hat{p})^2}
-\frac{2}{(\bar{p})^2}-\frac{1}{(p)^2}\right)+M^{IJ}M_{IJ}\frac{\left(p^0\right)^2p^2}{2(\bar{p})^2 (p)^2}\cr
&&-\left(s M^{23}+M^{12} M^{31}\right)\frac{p^0p^1}{\hat{p}\, p\, \bar{p}}-\left(-sM^{31}+M^{12}M^{23}\right)\frac{p^0p^2p^3}{(\hat{p})^3\, p}\cr
&&+sM^{12}\frac{p^1p^3}{(\hat{p})^2 \bar{p}}
+M^{12}\frac{p^1\left(-p^0p^3t+p^Ip_I z\right)}{(\hat{p})^2 \bar{p}}
+M^{23}\frac{p^1\, p\, t}{\hat{p}\bar{p}}\cr
&&-M^{31}\frac{p^2p^3\, p\, t}{\hat{p}(\bar{p})^2}
\end{eqnarray}
where $p=\sqrt{-p_\mu p^\mu}$, $\hat p=\sqrt{p_i p^i}$, $\bar p=\sqrt{p_I p^I}$. $\left(y^1,y^2,y^3\right)$ are constrained on unit sphere $S^2$ so that $\left(\theta,\phi\right)$ are coordinates of the $S^2$ and corresponding conjugate momenta are $\left(k^\theta,k^\phi\right)$. Moreover, $M^{IJ}$ are the angular momentum on the $S^2$. i.e.
\begin{eqnarray}
M^{12}&=&k^\phi\\
M^{23}&=&-k^\phi \cos\phi\cot\theta-k^\theta \sin\phi\\
M^{31}&=&-k^\phi \sin\phi\cot\theta+k^\theta \cos\phi
\end{eqnarray}
They indeed satisfy $SO\left(2,3\right)$ algebra.

In summary we have seen that in higher spin gauge theory one can effectively solve  the gauge fixing conditions (for Fronsdal fields in particular) and obtain the reduced physical set of fields and equation. 
The physical degrees of freedom of all spins were shown to be collected into a $6$-dimensional unconstrained scalar field, the six dimensions consisting of AdS$_4$ and S$^2$.  
This is the same number of degrees of freedom contained in the bi-local field derived from CFT.
Consequently at the level of unconstrained physical fields we can make a one-to-one identification between collective and reduced Higher Spin degrees of freedom. For this Map one only needs to give the change of coordinates (and momenta). Such canonical transformations were constructed in \cite{Koch:2010cy}.

In the final section we will return to the gauge world line particle framework and demonstrate the equivalence relations between the constraint algebras, giving another scheme for the Map.

\section{Equivalences}

In this section we will demonstrate that the algebras of the constraints specifying various  gauges match.
At the algebraic level the gauges are defined  by the structure constants of these algebras.
Consequently, if the algebra of the constraints can be seen to match in a particular basis equivalence between the
two gauges follows.

\subsection{Isomorphism of the KLSS and Collective Gauges}

The constraints specifying the KLSS gauge are given by the $\chi$'s below, while the constraints specifying the collective gauge are given by the $\eta$'s. Labeling the constraints as
\begin{eqnarray}
\chi_A=\frac{1}{2}(-X\cdot P + Y\cdot K)\qquad &&\eta_A=\frac{1}{2}(u\cdot P_u+v\cdot P_v)\cr
\chi_B=\frac{1}{2}(X\cdot P + Y\cdot K) \qquad &&\eta_B=\frac{1}{2} u\cdot P_u\cr
\chi_C=K \cdot K\qquad &&\eta_C= P_u\cdot P_u\cr
\chi_D=X \cdot K\qquad &&\eta_D= P_v\cdot P_v
\end{eqnarray}
we obtain the following Lie algebras
\begin{eqnarray}
\big[\chi_A,\chi_B\big]=0\qquad& \big[\eta_A,\eta_B\big]=0\cr
\big[\chi_C,\chi_D\big]=0\qquad &\big[\eta_C,\eta_D\big]=0\cr
\big[\chi_A,\chi_C\big]=-\chi_C\qquad &\big[\eta_A,\eta_C\big]=-\eta_C\cr
\big[\chi_A,\chi_D\big]=-\chi_D\qquad &\big[\eta_A,\eta_D\big]=-\eta_D\cr
\big[\chi_B,\chi_C\big]=-\chi_C\qquad &\big[\eta_B,\eta_C\big]=-\eta_C\cr
\big[\chi_B,\chi_D\big]=0\qquad &\big[\eta_B,\eta_D\big]=0
\end{eqnarray}
There is a complete match between the two algebras. 

\subsection{Equality of the Fronsdal and Collective Gauges}

For the collective (bilocal) description impose the following (first class) constraints
\begin{eqnarray}
U^2=0\qquad &V^2=0\cr
U\cdot P_U=0\qquad &V\cdot P_V=0
\end{eqnarray}
To obtain the correct Laplacian
\begin{equation}
\Omega_2+{1\over 4}\Omega_1^2 \longrightarrow P_U^2 P_V^2
\end{equation}
impose the little gauge condition
\begin{equation}
U\cdot V=1
\end{equation}

For the Fronsdal description, impose the following (again first class) constraints
\begin{eqnarray}
X\cdot P+Y\cdot K =0\qquad &X\cdot K=0\cr
P\cdot K=0\qquad &K\cdot K=0
\end{eqnarray}
To obtain the correct Laplacian
\begin{equation}
\Omega_2+{1\over 4}\Omega_1^2 \longrightarrow P^2
\end{equation}
impose the little gauge conditions
\begin{equation}
X\cdot Y=0\qquad X^2=1
\end{equation}

To map between these two sets of constraints, start by setting
\begin{eqnarray}
X=U+V\qquad &Y=U-V\cr
U=\frac{1}{2} (X+Y)\qquad &V=\frac{1}{2} (X-Y)\cr
P_U=P+K\qquad &P_V=P-K
\end{eqnarray}
Apply this change of coordinates to the collective constraints. The constraints $U^2=0=V^2$ becomes
\begin{equation}
X^2+Y^2=0\qquad X\cdot Y=0
\end{equation}
$PP_U+VP_V=0$ becomes
\begin{equation}
X\cdot P+Y\cdot K=0
\end{equation}
and $PP_U-VP_V=0$ becomes
\begin{equation}
X\cdot K +Y\cdot P =0
\end{equation}
Now perform the canonical transformation $Y\to K$ and $K\to -Y$. The collective constraints become
\begin{eqnarray}
X\cdot P -Y\cdot K=0\qquad &X\cdot K=0\cr
X^2+K^2=0\qquad &-X\cdot Y+P\cdot K=0
\end{eqnarray}
Now add the little gauge conditions
\begin{equation}
X^2=1\qquad X\cdot Y=0
\end{equation}
so that the constraints become
\begin{eqnarray}
X\cdot P -Y\cdot K=0\qquad &X\cdot K=0\cr
1+K^2=0\qquad &P\cdot K=0
\end{eqnarray}
There are two unfamiliar features: first $K^2=-1$ so that $K$ is a set of coordinates for de Sitter space.
Second, $X\cdot P +Y\cdot K$ is now $X\cdot P -Y\cdot K$. To understand the second point, note that Fronsdal uses
\begin{equation}
h_{A_1\cdots A_s}Y^{A_1}\cdots Y^{A_s}
\end{equation}
Acting on these fields, we have
\begin{equation}
X\cdot P +Y\cdot K\quad\longrightarrow\quad X\cdot{\partial\over\partial X}+Y\cdot{\partial\over\partial Y}
\end{equation}
For collective we use
\begin{equation}
h_{A_1\cdots A_s}K^{A_1}\cdots K^{A_s}
\end{equation}
Acting on these fields, we have (ignore ordering issues)
\begin{equation}
X\cdot P -K\cdot Y\quad\longrightarrow\quad X\cdot{\partial\over\partial X}+K\cdot{\partial\over\partial K}
\end{equation}
so there is again a perfect match.

\section{Conclusions}

We have presented two (equivalent) schemes for demonstrating the comparison between higher spin fields 
and bi-local fields built in CFT. In the first, the $3+3$ dimensional bi-local field equation is up-lifted to a $5+5$ dimensional  gauged system. It is seen that it represents a symmetric gauge of this HS theory in its world line description. It is then compared with a similar realization of Fronsdal's Higher Spin equations and also the Higher Spin particle system of \cite{Kuzenko:1994vh}. Equivalences between the gauges were then established at the level of constraints, namely, we have seen that they represent isomorphic algebras.
For direct comparison of bulk fields and equations  we have proceeded to solve the gauge constraints imposed on the Higher Spins and obtain equations entirely in terms of independent (gauge invariant) variables with no constraints. We were  able to perform this reduction in the Higher Spin case and show that it leads to a unconstrained scalar in $6$ dimensions, with space-time given by a product of AdS$_4$ and a $2$-sphere $S^2$ representing the reduced spin degrees of freedom. Collective fields mimic the dynamics of these unconstrained (invariant) fields. This reduction is a Higher Spin analog of a reduction to invariant fields known in pure gravity \cite{Bardeen:1980kt,Mukhanov:1990me}.

The world line scheme that we have employed for the present considerations  most directly concerns the quadratic level of the 
theory and also the one loop equivalences seen in direct calculations  performed in\cite{Giombi:2013fka,Jevicki:2014mfa}
(see also \cite{Strassler:1992zr}).
The structure of gauges that  we identify should in principle extend to the interacting case. 
One could hope that much like for strings the world line picture of spinning particles can be extended to the case of  
n-point amplitudes.  
The fully nonlinear collective action indeed features  bilocal  Feynman rules characteristic of a spinning dipole. 
This direction is worthy of future study.

The present scheme of establishing the correspondence gives a different perspective of holography in AdS/CFT as compared to the standard projections to the boundary. Usually the correspondence is established through comparison of boundary correlation functions. In the present discussion we are establishing the correspondence in the  bulk and off-shell. It is seen that collective field theory can be identified with the reduced, unconstrained fields of Higher Spin Gravity. It therefore summarizes the physical, `gauge invariant' data of the theory. We should mention that there are other, possibly related views on the holography in Higher Spin theories. In the renormalization group construction of \cite{Leigh:2014qca}
bi-local observables are extended by additional (gauge) degrees of freedom. Comparison regarding the origin of AdS space is of interest. At the level of gauge invariant Higher Spin equation on has the proposals of Vasiliev \cite{Vasiliev:2012vf,Vasiliev:2014vwa}. Relationships between all these approaches to Holography 
are of major interest.

{\vskip 0.5cm}

\noindent
{\it Acknowledgements:}
The work of AJ and JY is supported by the Department of Energy under contract DE-FG-02-91ER40688. AJ also acknowledges the hospitality of National Center for Theoretical Physics,University of Wittwatersrand where this work was done. RdMK is supported by the South African Research Chairs Initiative of the Department of Science and Technology and National Research Foundation.
The work of JP Rodrigues is based on the research supported in part by
the National Research Foundation of South Africa (Grant specific unique reference number (UID 85974).

\end{document}